\documentclass[aps,prd,superscriptaddress,floatfix,showpacs]{revtex4}

\usepackage{verbatim}
\usepackage[utf8]{inputenc}

\usepackage{amsmath}
\usepackage{amsfonts}
\usepackage{amssymb}
\usepackage{graphicx}
\usepackage{bm}
\usepackage{color}
\usepackage{epsf}

\usepackage{psfrag}

\newcommand{\beq}[1]{
\begin{equation}\label{#1}}
\newcommand{\eeq}{\end{equation}}
\newcommand{\bea}[1]{
\marginpar{\small\textsf{#1}}
\begin{eqnarray}\label{#1}}
\newcommand{\eea}{\end{eqnarray}}
\def\bea{\begin{eqnarray}}
\def\eea{\end{eqnarray}}
\def\beas{\begin{eqnarray*}}
\def\eeas{\end{eqnarray*}}
\def\beqas{\begin{eqnarray*}}
\def\eqas{\end{eqnarray*}}
\def\beq{\begin{equation}}
\def\eeq{\end{equation}}
\def\beqd{\begin{displaymath}}
\def\eeqd{\end{displaymath}}
\def\eqd{\end{displaymath}}

\def\slashchar#1{\setbox0=\hbox{$#1$}
   \dimen0=\wd0
   \setbox1=\hbox{/} \dimen1=\wd1
   \ifdim\dimen0>\dimen1
      \rlap{\hbox to \dimen0{\hfil/\hfil}}
      #1
   \else\begin{eqnarray}
      \rlap{\hbox to \dimen1{\hfil$#1$\hfil}}
      /
   \fi}

\begin{document}
\title
{Charged current electroproduction of a charmed meson at an electron-ion collider}
\author{ B.~Pire}
\affiliation{ Centre de Physique Th\'eorique, CNRS, \'Ecole Polytechnique,
 I.P. Paris, 91128 Palaiseau,     France }

\author{ L.~Szymanowski}
\affiliation{ National Centre for Nuclear Research (NCBJ), 02-093 Warsaw, Poland}

\author{  J. Wagner}
\affiliation{ National Centre for Nuclear Research (NCBJ), 02-093 Warsaw, Poland}
\date{\today}
\begin{abstract}

\noindent
We calculate the amplitude for exclusive  electroweak production of a pseudoscalar  $D_s$ or a vector $D^*_s$  charmed strange meson on an unpolarized nucleon, through a charged current, in leading order in $\alpha_s$. We work in the framework of the collinear QCD approach where generalized gluon distributions factorize from  perturbatively calculable coefficient functions. We include both $O(m_c)$ terms in the coefficient functions and $O(M_D)$ mass term contributions in the heavy meson distribution amplitudes. We show that this process may be accessed at future electron-ion colliders.\end{abstract}

\maketitle

\section{Introduction.}

Exclusive electroproduction processes involving charged currents (i.e. through a $W^\pm$ exchange) have not been much discussed up to now, with the notable exception of the pioneering work \cite{Siddikov:2019ahb}. The reason is simple as the smallness of the weak coupling prevents exclusive cross-sections from being large enough to allow sufficient counting rates at existing electron-nucleon facilities. The very high luminosity anticipated at planned high energy electron-ion colliders \cite{Accardi:2012qut,Anderle:2021wcy} should open this physics domain to a detailed investigation of various interesting channels. In this respect, the production of a single charmed meson - which is forbidden in pure electromagnetic processes - is a specific way to study various features of hadronic physics, and in particular effects of the heavy quarks mass in the framework of collinear QCD factorization.  
Indeed, the well established framework of collinear QCD factorization \cite{fact1,fact2,fact3} for scattering amplitudes in exclusive electroproduction, in reactions mediated by a highly virtual photon, may also be applied to reactions mediated by a virtual $W^\pm$ boson,  in a similar generalized Bjorken regime \cite{weakGPD}. This framework describes hadronic amplitudes using generalized parton distributions (GPDs) which give access to a 3-dimensional analysis \cite{3d} of  the internal structure of hadrons. 

Since  charged currents are mediated by a massive vector boson exchange which is usually highly virtual, one is  tempted to apply a factorized description of the process amplitude down to quite small values of the momentum transfer $Q^2=-q^2$ carried by the $W^\pm$ boson. Moreover, heavy quark production allows to extend the range of validity of collinear factorization, the heavy quark mass playing the role of the hard scale. Indeed kinematics (detailed below) shows that the relevant scale is $O(Q^2+m_c^2)$.

\begin{figure}
\begin{center}
\includegraphics[width=0.9\textwidth]{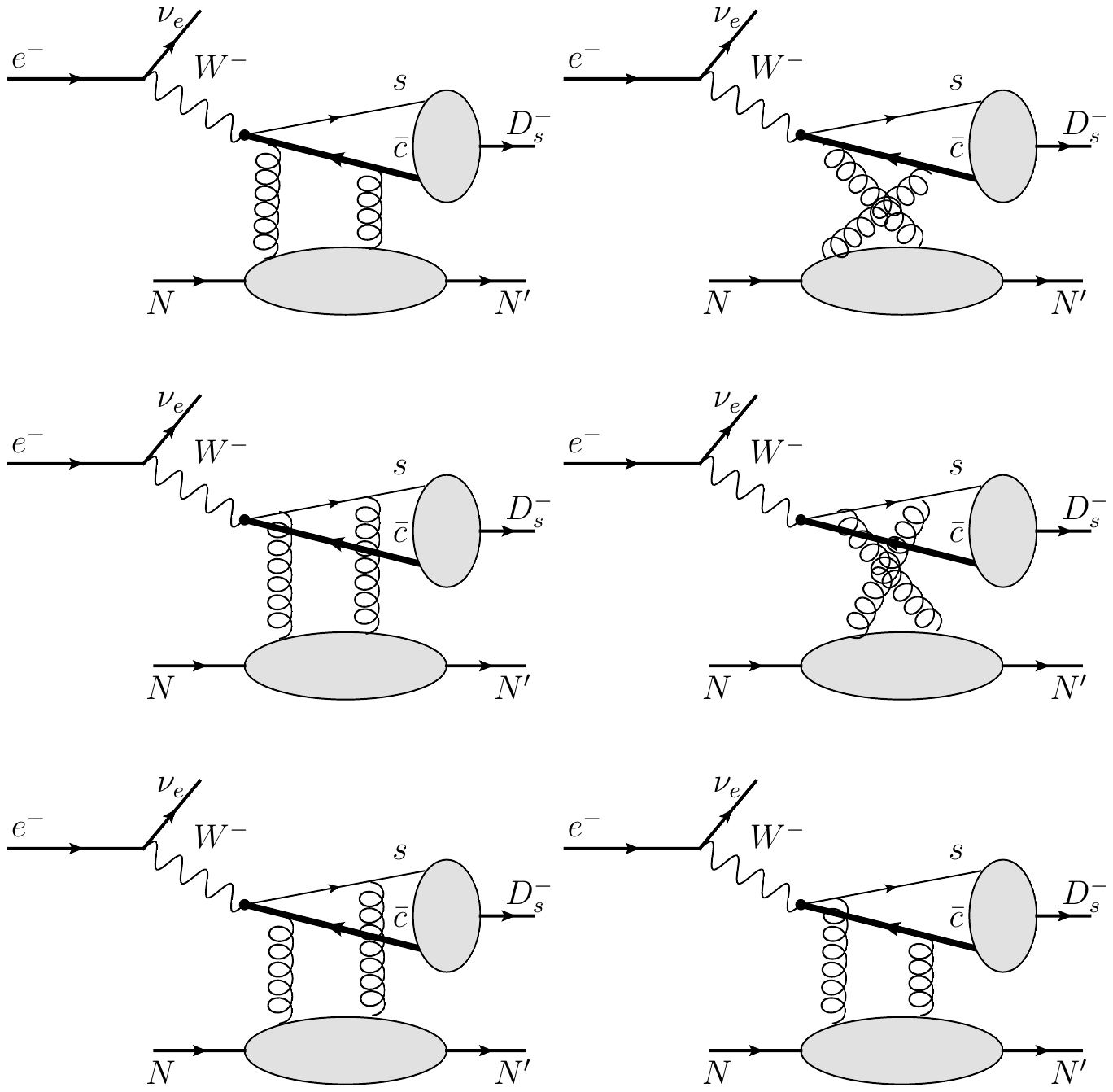}
\end{center}
\vspace{-9cm}
\caption{Feynman diagrams for the factorized  amplitude for the $e^- +N \to \nu_e + D_s^{-} + N'$ process involving the gluon GPDs; the thick line represents the heavy anti-quark $\bar c$.}
   \label{Fig1}
\end{figure}

Since the scattering amplitude is proportional to the relevant CKM matrix element, the dominant production process for  a charmed meson  involves a $D^-_s(1968)$ or  $D^{*-}_s(2112)$ charmed and strange meson. The $D^{*-}_s(2112)$ decays mostly in a $D^-_s(1968) \gamma$ pair.
In this paper we shall thus restrict our study to the exclusive production of a pseudoscalar $D_s^-$ or a   $D^{*-}_s$ vector meson through the reactions on a nucleon N (proton or  neutron) target:
\begin{eqnarray}
e^- (k)+N(p_1) &\to& \nu_e (k')+D_s^- (p_D)+N'(p_2) \,,\\
e^- (k)+N(p_1) &\to& \nu_e (k')+D_s^{*-} (p_D)+N'(p_2)\,,
\end{eqnarray}
in the kinematical domain where collinear factorization  leads to a description of the scattering amplitude as a convolution of gluon GPDs and the $D_s^-$ or $D_s^{*-}$ meson distribution amplitude (DA) (see Fig.\ref{Fig1}) with
 the amplitude for hard subprocesses:
\begin{eqnarray}
W^- + g \to (\bar c~s) +g\,,
\end{eqnarray}
 calculated in the collinear kinematics taking heavy quark mass effects into account \cite{PS,Pire:2017lfj}. In order to be consistent, we shall include the order $\frac{M_D}{Q^2+M_D^2}$  contributions related to mass terms in the distribution amplitudes of heavy mesons (see Eq. (\ref{DA})).

Our kinematical notations are as follows ($m$ and $M_D$ are the nucleon and $D_s-$meson masses):
\begin{eqnarray}
&&q=k-k'~~~~~; ~~~~~Q^2 = -q^2~~~~~; ~~~~~\Delta = p_2-p_1 ~~~~~; ~~~~~\Delta^2=t \,;\nonumber\\
&&q^\mu= -2\xi' p^\mu +\frac{Q^2}{4\xi'} n^\mu ~;~\epsilon_L^\mu(q)= \frac{1}{Q} [2\xi' p^\mu +\frac{Q^2}{4\xi'} n^\mu] ~;~p_D^\mu=  2(\xi-\xi') p^\mu +\frac{M_D^2-\Delta_T^2}{4(\xi-\xi')}  n^\mu -\Delta_T^\mu \,; \\
&&p_1^\mu=(1+\xi)p^\mu +\frac{1}{2}  \frac{m^2-\Delta_T^2/4}{1+\xi} n^\mu -\frac{\Delta_T^\mu}{2}~~~~;~~~~ p_2^\mu=(1-\xi)p^\mu +\frac{1}{2}  \frac{m^2-\Delta_T^2/4}{1-\xi} n^\mu +\frac{\Delta_T^\mu}{2}
 \,,\nonumber
\end{eqnarray}
with $p^2 = n^2 = 0$ and $p \cdot n=1$. 
We define the skewness variable as :
 \begin{eqnarray}
\xi = - \frac{(p_2-p_1)\cdot n}{2}  \,;
\label{skewness}
\end{eqnarray} 
neglecting the nucleon mass and $\Delta_T$, its approximate value is
\begin{eqnarray}
\xi \approx \frac{Q^2+M_D^2}{4p_1\cdot q-Q^2-M_D^2}  \,,
\end{eqnarray}
while $\xi'= -q \cdot n /2 = \frac{\xi Q^2}{Q^2+M_D^2}$.

To unify the description of the scaling amplitude, we thus define a modified Bjorken variable
 \begin{eqnarray}
x_B^D \equiv \frac {Q^2+M_D^2}{2p_1  \cdot q} \neq x_B \equiv \frac {Q^2}{2p_1 \cdot q}\,,
\end{eqnarray}
which allows to express $\xi$ in a compact form:
 \begin{eqnarray}
\xi \approx \frac{x_B^D }{2-x_B^D } \,.
\end{eqnarray}
If the meson mass is the relevant large scale (for instance in the limiting case where $Q^2$ vanishes as in the timelike Compton scattering kinematics \cite{TCS}) :
\begin{eqnarray}
 \xi \approx \frac{\tau}{2-\tau} ~~;~~\tau = \frac{M_D^2}{s_{WN}-m^2}\,.
\end{eqnarray}

\section{Distribution amplitudes and GPDs} 
\begin{figure}
\includegraphics[width=0.6 \textwidth]{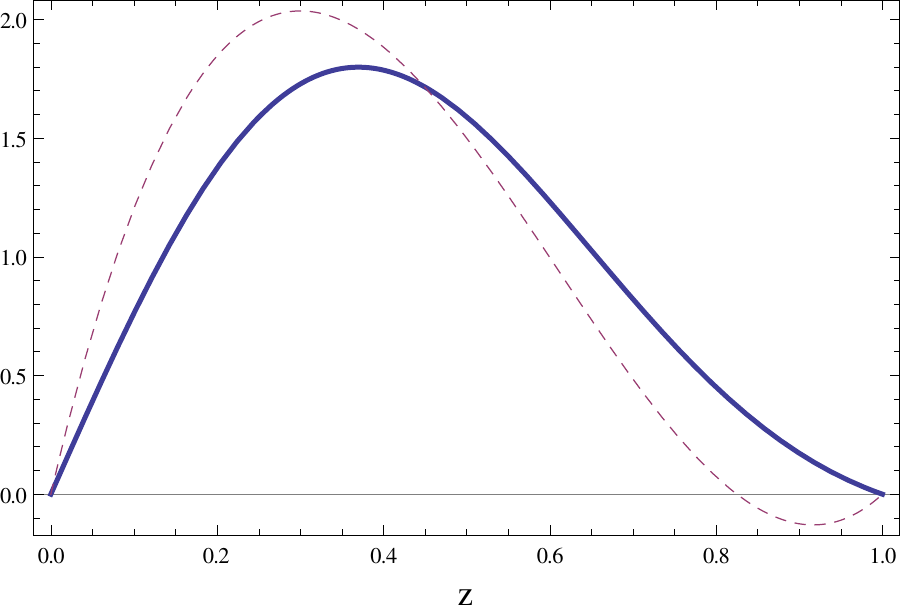}
\caption{The $D_s$ meson distribution amplitudes proposed  by Ref.\cite{heavyDA2} (dashed curve) and by  Ref.\cite{Serna:2020txe} (solid curve).} 
   \label{DAs}  
\end{figure}
In the collinear factorization framework, the hadronization of the quark-antiquark pair is described by a distribution amplitude (DA) which obeys a twist expansion and evolution equations. Much work has been devoted to this subject \cite{heavyDA}.
The charmed meson distribution amplitudes are less known than the light meson ones.  Here, we shall 
follow Ref. \cite{heavyDA2} and include some mass terms which will lead to order $\frac{M_D}{Q^2+M_D^2}$ contributions to the amplitudes;   omitting the path-ordered gauge link, the relevant distribution amplitude reads for the pseudo scalar $D_s^-$ meson:
\begin{eqnarray}
\langle D^-_s(P_D) | \bar s_\beta(y) c_\gamma(-y) |0 \rangle & =&
   i \frac{f_{D_s}}{4} \int_0^1 dz e^{i(z-\bar z)P_D\cdot y} [(\hat P_D-M_D)\gamma^5]_{\gamma \beta}   \phi_{D_s}(z)\,,
   \label{DA}
         \end{eqnarray}
where $\int_0^1 dz ~ \phi_{D_s}(z) = 1$. As usual, we denote $\bar z=1-z$ and  $\hat p = p_\mu \gamma^\mu$ for any vector $p$. 
 Contrarily to the case of light mesons, where DAs are symmetric in a $z \to \bar z$ transformation, heavy meson DAs are asymmetric, the heavy quark (or antiquark) taking most of the light cone momentum of the hadron. In our case of a heavy antiquark-light quark meson, this means a DA which is strongly peaked around $z_0$ with $1-z_0 = \frac {m_c}{M_D}$. 
 In our estimates, we will thus parametrize  $\phi_{D_s}(z) $ following two recent studies:
 \begin{itemize}
     \item as in Ref. \cite{heavyDA2}, {\it i.e.} $\phi_{D_s}(z) =6z(1-z)(1+C_D(1-2z))$ with $C_D\approx1.5$, which has a maximum around $z=0.3$,
     \item as in Ref. \cite{Serna:2020txe}, {\it i.e.} $\phi_{D_s}(z) =3.12z(1-z)e^{2.85 z(1-z)-0.93(2z-1)}$\,.
 \end{itemize}
As shown in Fig.\ref{DAs}, these two parametrizations are somewhat different, and will allow us to roughly quantify the uncertainty of our estimates with respect to a reasonable choice of the $D_s$'s distribution amplitude.  
The coupling constant $f_{D_s}$ of the  pseudoscalar meson $D_s$ has recently been calculated on the lattice as $f_{D_s}\approx  0.24$ GeV \cite{Blossier:2018jol,Aoki:2019cca}.

The DA of the vector meson $D^*_s$ is defined as in the case of the  $D^*$ vector meson case, and for the  longitudinal and transverse polarization states read:
\begin{eqnarray}
\langle D_s^{*-}(P_D, \varepsilon_L) | \bar s_\beta(y) c_\gamma(-y) |0 \rangle & =&
  \frac{f_{D^*_s}}{4} \int_0^1 dz e^{i( z- \bar z)P_D.y} [\hat P_D - M_D]_{\gamma \beta}   \phi_{D^*_s}(z)\,,
\\
\langle D^{*-}_s(P_D,\varepsilon_T) | \bar s_\beta(y) c_\gamma(-y) |0 \rangle & =&
   \frac{f_{D^*_s}}{4} \int_0^1 dz e^{i(z-\bar z)P_D.y} [ (\hat P_D - M_D)\hat \varepsilon_T]_{\gamma \beta}   \phi_{D^*_s}(z)\,,
         \end{eqnarray}
         where $\int_0^1 dz ~ \phi_{D^*_s}(z) = 1$.
The coupling constants $f_{D^*_s}$ may be different for the transversally and longitudinally polarized mesons, but the difference is unlikely to be large.
The ratio of the coupling constants of the $D_s^*$ vector meson to the $D_s$ pseudoscalar meson has been estimated on the lattice \cite{Blossier:2018jol,Lubicz:2017asp} as $f_{D^*_s}/ f_{D_s} \approx 1.1$. Since no parametrization of the DA of the $D_s^*$ vector meson has been proposed up to now, we shall use the same shape as for the pseudoscalar meson, $\phi_{D_s}(z)=\phi_{D^*_s}(z)$ .

We define the  generalized parton distributions of a parton  in the nucleon target   with the conventions of \cite{MD}.
To get the quantitative predictions, we use the Goloskokov-Kroll (G-K) model for gluonic GPDs, based on the fits to deeply virtual meson production. Details of the model can be found in \cite{CEGPD}. To quantify the dependence of our predictions on the poorly known gluon distributions, we present also the results based on the other simple model of GPDs described in \cite{Moutarde:2013qs}.

 \section{The $D_s$ meson production amplitude.}
   If we neglect the strange quark content of the nucleon, there is no contribution coming from quark GPDs and the only relevant  contribution comes from the diagrams of Fig. \ref{Fig1} with the gluon GPDs. The expression for the amplitudes can be read off our previous work \cite{Pire:2017lfj} on neutrino-production. 
   For completeness, we copy the relevant equations (with appropriate exchange of $z$ and $\bar{z}$), neglecting the strange quark mass.

The six Feynman diagrams of Fig. \ref{Fig1} contribute to the coefficient function. The  last three ones correspond to the  first three ones with the substitution $x\leftrightarrow -x$, and an overall minus sign for the axial case.
The transversity gluon GPDs do not contribute to the longitudinal amplitude since there there is no way to flip the helicity by two units when producing a (pseudo)scalar meson. This will not be the case for the production of a vector meson $D_s^*$.

The symmetric and antisymmetric hard amplitudes read:
\begin{eqnarray}
g_\perp^{ij} {\cal M}_H^S &=&   \left\{ \frac{Tr_a^S}{D_1 D_2} + \frac{Tr_b^S}{D_3 D_4} + \frac{Tr_c^S}{D_4 D_5}\right\}  + \left\{x\rightarrow  -x\right\}\,, \\
i \epsilon_\perp^{ij}{\cal M}_H^{A} &=&  \left\{ \frac{Tr_a^A}{D_1 D_2} + \frac{Tr_b^A}{D_3 D_4} + \frac{Tr_c^A}{D_4 D_5} \right\}- \left\{x\rightarrow  -x\right\}\,,
\end{eqnarray}
where the traces are 
\begin{eqnarray}
Tr_a^S &=&  \frac{2  z }{Q} g_T^{ij}\left[\bar z M_D^4 +Q^4+Q^2M_D^2(2- z) - \frac{x+\xi}{2\xi} Q^2 (Q^2+M_D^2)\right]\,,\\
Tr_a^A &=&  \frac{2i  z \epsilon^{npij}}{Q} \left[\bar z M_D^4 +Q^2M_D^2(1+ z) + \frac{x-\xi}{2\xi} Q^2 (Q^2+M_D^2)\right]\,,\\
Tr_b^S &=&   \frac{2(Q^2+M_D^2) }{Q} g_T^{ij}\left[- \frac{x+\xi}{2\xi} m_cM_D - \bar z \frac{x-\xi}{2\xi}Q^2 +m_c^2+z \bar z M_D^2 \right]\,,\\
Tr_b^A &=&  \frac{2i  \epsilon^{npij}}{Q} \left[-\bar z\frac{x-\xi}{2\xi} Q^4 +Q^2M_D^2(2-z)  (-\bar z- \frac{x-\xi}{2\xi})+ M_D^4(\bar z^2-\bar z-1- \frac{x+\xi}{2\xi})\right.\nonumber \\
&+& \left. (M_D^2-m_c^2) (M_D^2-Q^2) +M_D(M_D+m_c)(Q^2+M_D^2)\frac{x+\xi}{2\xi} \right]\,,\\
Tr_c^S &=&  -\frac{Q^2+M_D^2}{\xi Q}g_T^{ij} \left[ (Q^2+M_D^2) \frac{x^2-\xi^2}{2\xi} +2 \bar z M_D^2(\xi \bar z+x) -M_D(m_c+M_D)(x-\xi+2\xi \bar z)         \right] \,,\\
Tr_c^A &=&   \frac{2i  \epsilon^{npij}}{Q} \left[(\bar z^2-1)Q^2M_D^2 +( z M_D^2-(Q^2+M_D^2)\frac{x+\xi}{2\xi})((2- z) M_D^2+(Q^2+M_D^2)\frac{x-\xi}{2\xi})\right.\nonumber\\ 
&+& \left.M_D(m_c+M_D)[(Q^2+M_D^2)\frac{x+\xi}{2\xi} + z(Q^2-M_D^2)]\right]\,,
\end{eqnarray}
and the denominators read (with $\alpha = \frac{2\xi M_D^2}{M_D^2+Q^2}$):
\begin{eqnarray}
D_1 &=&  z [  -\bar z M_D^2 -Q^2 +i \varepsilon] \,,\\
D_2 &=&  z [  z M_D^2 +\frac{x-\xi}{2\xi}(Q^2+M_D^2) +i \varepsilon] =   z \frac{Q^2+M_D^2}{2\xi} (x-\xi+\alpha z +i\epsilon)\,,\\
D_3 &=& -\bar z Q^2-z \bar z M_D^2- m_c^2 +i \varepsilon = -\bar z(Q^2+M_D^2) +\bar z^2M_D^2-m_c^2 +i\epsilon \,,\\
D_4 &=& \bar z^2 M_D^2 -m_c^2+\frac{\bar z(x-\xi)}{2\xi}(Q^2+M_D^2) +i \varepsilon\,,\\
D_5 &=&  z [  z M_D^2 -\frac{x+\xi}{2\xi}(Q^2+M_D^2) +i \varepsilon]  =  z \frac{Q^2+M_D^2}{2\xi} (-x-\xi+\alpha z +i\epsilon)\,.
\end{eqnarray}
The gluonic contribution to the amplitude thus reads:
\begin{eqnarray}
T_L^g &=&  \frac{ i C_g}{2} \int_{-1}^{1}dx \frac{- 1}{(x+\xi-i\epsilon)(x-\xi+i\epsilon)} \int_0^1 dz f_{D_s} \phi_{D_s}(z) \cdot \nonumber \\
&& \left[  \bar{N}(p_{2})[H^g\hat n +E^g\frac{i\sigma^{n\Delta}}{2m} ]N(p_{1}) {\cal M}_H^S+\bar{N}(p_{2})[{\tilde H}^g \hat n \gamma^5+{\tilde E}^g\frac{\gamma^5n \cdot \Delta}{2m} ]N(p_{1}) {\cal M}_H^{A}  \right] \, \\
&\equiv&  \frac{- i C_g}{2Q}  \bar{N}(p_{2}) \left[ {\cal H}^g\hat n +{\cal E}^g\frac{i\sigma^{n\Delta}}{2m} +{\tilde {\cal H}}^g \hat n \gamma^5+{\tilde {\cal E}}^g\frac{\gamma^5n\cdot\Delta}{2m} \right] N(p_{1}) \,, 
\end{eqnarray}
where  the last line defines the gluonic form factors 
${\cal H}^g$, $\tilde {\cal H}^g$, ${\cal E}^g$, $\tilde {\cal E}^g$ and  $C_g= T_f\frac{\pi}{3} \alpha_s V_{sc}$ 
with $T_f=\frac{1}{2}$ and the factor $\frac{- 1}{(x+\xi-i\epsilon)(x-\xi+i\epsilon)}$ comes from the conversion of the strength tensor to the  gluon field. Note that there is no singularity in the integral over $z$ if the DA vanishes like $z \bar z$ at the limits of integration.


 \section{The $D^*_s$ meson production amplitude.}
 
 Let us now consider  the exclusive production of the  $D_s^{*-}(2112)$ vector meson through the reaction:
\begin{eqnarray}
e^-(k) + p(p_1, \lambda) &\to& \nu_e(k')+ D_s^{*-} (p_D,\varepsilon_D)+p'(p_2, \lambda') .
\end{eqnarray}
In \cite{Pire:2017yge}, we showed that neutrino production of $D^*$ vector mesons may help to measure the gluon transversity GPDs, the phenomenology of which is presently restricted to angular asymmetries in DVCS  \cite{Belitsky:2000jk}, which turns out to be quite difficult to access experimentally. We do not follow this task here since the reconstruction of the decay products of the $D_s^*$ - which must be carried on to isolate the transversity gluon GPD contribution - is likely to be a too hard challenge. The amplitude for charged current $D_s^{*-}$ production may be read off  from the neutrino production study \cite{Pire:2017yge}. There are three non-zero ($W^- \to D_s^{*-}$) helicity amplitudes:
\begin{itemize}
\item a longitudinal ($W^-$) to longitudinal ($D_s^{*-}$) amplitude ${\cal M}_{00}$;
\item a left ($W^-$) to left ($D_s^{*-}$)   ${\cal M}_{LL}$;
\item a left ($W^-$) to right ($D_s^{*-}$)  ${\cal M}_{LR}$ , which is proportional to transversity gluon GPDs.
\end{itemize}

 Apart from trivial changes in the masses and coupling constants, the amplitude for the longitudinally polarized vector meson $D_s^*$ production is calculated in the same way as the one for the pseudoscalar $D_s$ production. The additional $\gamma_5$ matrix in the definition of the pseudoscalar DA does not alter the magnitude of the coefficient function acting on the gluon GPD, as previously shown in the massless quark case of $\pi$ vs $\rho_L$ production \cite{Pire:2017tvv}; this result is also true for the massive charm quark case studies here. The main difference in the production rates will thus come from the higher values of the skewness $\xi$ at fixed values of $y$ and beam energy, which results in smaller values of the gluon GPDs.
 
 The amplitude for the transversely polarized $D_s^*$ production has two components; one which depends on the usual gluon GPDs and contributes to $\sigma_{--}$ and another one which depends on the gluon transversity GPD, the magnitude of which is unknown up to now, but does not contribute to the angular integrated cross-section.
The amplitude  ${\cal M}_{LL}$ which contributes to the azimuthal angle integrated cross-section is expressed in terms of form factors ${\cal H}^g_{T}$, ${\cal E}^g_{T}$, $\tilde {\cal H}^g_{T}$, $\tilde {\cal E}^g_{T}$ as  :

\begin{eqnarray}
 {\cal M}_{LL} &=&  \frac{ i C_g}{2} \int_{-1}^{1}dx \frac{- 1}{(x+\xi-i\epsilon)(x-\xi+i\epsilon)} \int_0^1 dz f_T \phi_{D_s^*}(z) \cdot \nonumber \\
&& \left[  \bar{N}(p_{2})[H^g\hat n +E^g\frac{i\sigma^{n\Delta}}{2m} ]N(p_{1}) G_T+\bar{N}(p_{2})[{\tilde H}^g \hat n \gamma^5+{\tilde E}^g\frac{\gamma^5n\cdot\Delta}{2m} ]N(p_{1}) \tilde G_T  \right]  \\
&\equiv&  \frac{- i C_g}{2Q}  \bar{N}(p_{2}) \left[ {\cal H}^g_T\hat n +{\cal E}^g_T\frac{i\sigma^{n\Delta}}{2m} +{\tilde {\cal H}}^g_T \hat n \gamma^5+{\tilde {\cal E}}^g_T\frac{\gamma^5n\cdot\Delta}{2m} \right] N(p_{1})\, ,
\label{FFT}
\end{eqnarray}
 with $T_f=\frac{1}{2}$ and  $C_g= T_f\frac{\pi}{3} \alpha_s V_{sc}$ for $D_s^*$ production, and where    $G_T$ and $\tilde G_T$ factors read \cite{Pire:2017yge}:
 
\begin{eqnarray}
   G_T &=&
 \frac{- 8 M_D  z \epsilon_D^* \cdot
   \epsilon_W \left(\kappa 
   (x-3\xi )+M_D^2
    z\right) } { D_1(x,\xi) D_2(x,\xi)} + 
   \frac{ 8 i M_D  z \kappa (x+\xi))
   \epsilon ^{p n \epsilon_D^*
\epsilon_W }}{D_1(x,\xi) D_2 (x,\xi)} 
\nonumber \\
& &-8 \frac {\kappa  (\bar z M_D+m_c) (x-\xi) \epsilon_D^* \cdot \epsilon_W }{D_3(x,\xi) D_4(x,\xi)} + \{ x \to -x  \}\,,\\
\tilde G_{T}&=& 
   \frac{ 8 i M_D  z~ p\cdot \epsilon_W ~p\cdot \epsilon_D^*  \left(M_D^2 (\xi-x-2\xi \bar z)+4 \kappa~ \xi ~(x-  \xi) \right)  }{\kappa~ D_1(x,\xi) D_2 (x,\xi)} 
   \nonumber \\
& &
   + \frac{ -8 M_D  z \epsilon ^{np\epsilon_D^* \epsilon_W} \left(M_D^2  z +\kappa  (x-3 \xi)\right) } { D_1(x,\xi) D_2(x,\xi)}
   - \{ x \to -x  \} \,,
   \label{G_LT}
  \end{eqnarray}
with $\kappa = \frac{M_D^2+Q^2}{4 \xi}$.

 \section{Observables}
 
  The (initial electron spin averaged) differential cross section for the production of a pseudoscalar charmed meson is written, after azimuthal integration, as \cite{Arens}:
  \begin{eqnarray}
\label{csphi}
\frac{d\sigma(e^- N\to \nu N'D_s^-)}{dy\, dQ^2\, dt}
 =  \pi \bar\Gamma
\Bigl\{ ~\frac{1+ \sqrt{1-\varepsilon^2}}{2} \sigma_{- -}&+&\varepsilon\sigma_{00} \Bigr\}.
\end{eqnarray}
with $y= \frac{p \cdot q}{p\cdot k}$ , $Q^2 = x_B y (s-m^2)$, $\varepsilon \approx \frac{1-y}{1-y+y^2/2}$ and
\begin{equation}
\bar \Gamma = \frac{G_F^2}{(2 \pi)^4} \frac{1}{32y} \frac{1}{\sqrt{ 1+4x_B^2m^2/Q^2}}\frac{1}{(s-m^2)^2} \frac{Q^2}{1-\epsilon}\,, 
\end{equation}
where the ``cross-sections'' $\sigma_{lm}=\epsilon^{* \mu}_l W_{\mu \nu} \epsilon^\nu_m$ are product of  amplitudes for the process $ W^-(\epsilon_l) N\to D^-_s N' $, averaged  (summed) over the initial (final) hadron polarizations.

 For pseudoscalar $D^-_s$-meson production, $\sigma_{--}$ vanishes while the longitudinal cross section $\sigma_{00}$ is straightforwardly obtained by squaring the sum of the amplitudes $ T^g_{L}$; at zeroth order in $\Delta_T$, it reads  :
\begin{eqnarray}
&&\sigma_{00}|_{D^-_s} = \nonumber \\
&&\frac{1} { Q^2}\biggl\{[\, | C_g{\mathcal{H}}^{g}|^2 + | C_g\tilde{\mathcal{H}}^{g}|^2 ] (1-\xi^2) +\frac{\xi^4}{1-\xi^2} [\, |C_g\tilde{\mathcal{E}}^{g} |^2 +  |C_g {\mathcal{E}}^{g} |^2]  \nonumber    -2 \xi^2 {\mathcal R}e  [ C_g{\mathcal{H}}^{g}] [C_g {\mathcal{E}}^{g*}]  -2 \xi^2 {\mathcal R}e  [ C_g\tilde{\mathcal{H}}^{g}] [C_g \tilde{\mathcal{E}}^{g*}] \biggr\} .\, \nonumber\\
&&
\label{sig00}
\end{eqnarray}
For transversely polarized vector $D^{*-}_s$ meson production the cross sections are given by Eq. (\ref{sig00}) after appropriate replacement of form factors in Eq. (\ref{sig00}) by those defined in Eq.(\ref{FFT}):
\begin{eqnarray}
&&\sigma_{00}|_{D^{*-}_s} = \nonumber \\
&&\frac{1} { Q^2}\biggl\{[\, | C_g{\mathcal{H}}^{g}_L|^2 + | C_g\tilde{\mathcal{H}}^{g}_L|^2 ] (1-\xi^2) +\frac{\xi^4}{1-\xi^2} [\, |C_g\tilde{\mathcal{E}}^{g}_L |^2 +  |C_g {\mathcal{E}}^{g}_L |^2]  \nonumber    -2 \xi^2 {\mathcal R}e  [ C_g{\mathcal{H}}^{g}_L] [C_g {\mathcal{E}}^{g*}_L]  -2 \xi^2 {\mathcal R}e  [ C_g\tilde{\mathcal{H}}^{g}_L] [C_g \tilde{\mathcal{E}}^{g*}_L] \biggr\} .\, \nonumber\\
&&
\label{sig00D*}
\end{eqnarray}
\begin{eqnarray}
&&\sigma_{--}|_{D^{*-}_s} = \nonumber \\
&&\frac{1} { Q^2}\biggl\{[\, | C_g{\mathcal{H}}^{g}_T|^2 + | C_g\tilde{\mathcal{H}}^{g}_T|^2 ] (1-\xi^2) +\frac{\xi^4}{1-\xi^2} [\, |C_g\tilde{\mathcal{E}}^{g}_T |^2 +  |C_g {\mathcal{E}}^{g}_T |^2]  \nonumber    -2 \xi^2 {\mathcal R}e  [ C_g{\mathcal{H}}^{g}_T] [C_g {\mathcal{E}}^{g*}_T]  -2 \xi^2 {\mathcal R}e  [ C_g\tilde{\mathcal{H}}^{g}_T] [C_g \tilde{\mathcal{E}}^{g*}_T] \biggr\} .\, \nonumber\\
&&
\label{sig--D*}
\end{eqnarray}

\begin{figure}[h]
\includegraphics[width=1\textwidth]{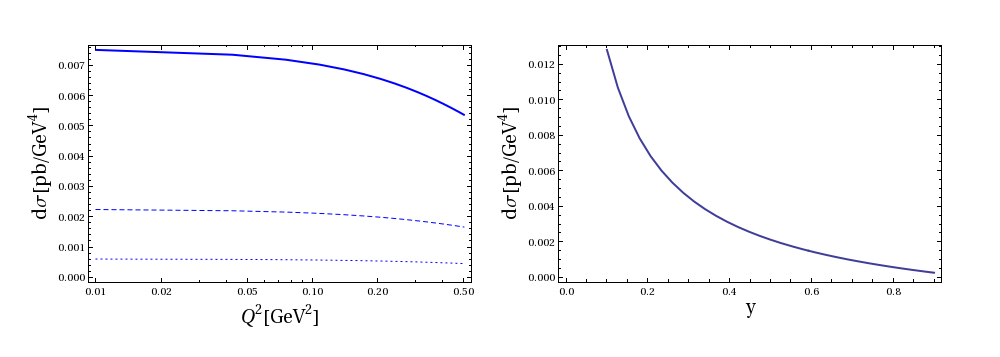}
\caption{Left panel : The $Q^2$ dependence  of  the cross section $\frac{d\sigma(e^- N \to \nu N D_s^-)}{dy\, dQ^2\, dt}$ (in pb GeV$^{-4}$) for  $\Delta_T = 0$  and $s=820$ GeV$^2$ and $y=.2$ (solid curve) , $y=.5$ (dashed curve) and $y=.8$  (dotted curve). Right  panel : The $y$ dependence of  the cross section $\frac{d\sigma(e^- N \to \nu N D_s^-)}{dy\, dQ^2\, dt}$ (in pb GeV$^{-4}$) for $Q^2=0.1$ GeV$^2$, $\Delta_T = 0$  and $s=820$ GeV$^2$.}
   \label{sigma_821}
\end{figure}

\begin{figure}[h]
\includegraphics[width=1\textwidth]{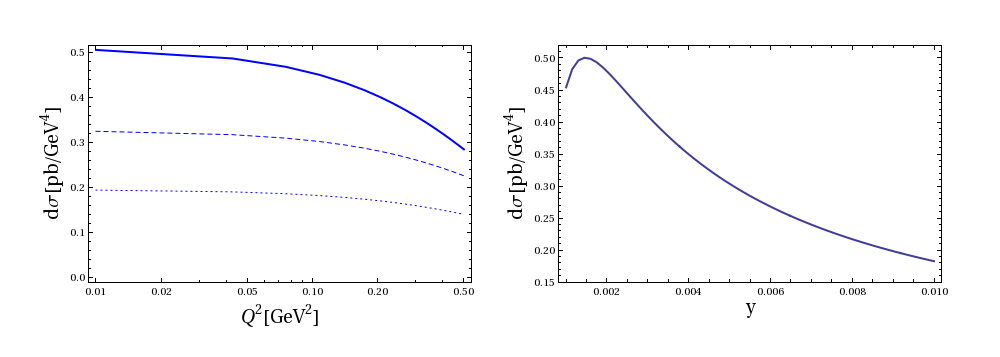}
\caption{Left panel : The $Q^2$ dependence  of  the cross section $\frac{d\sigma(e^- N \to \nu N D_s^-)}{dy\, dQ^2\, dt}$ (in pb GeV$^{-4}$) for  $\Delta_T = 0$  and $s=20000$ GeV$^2$ and $y=10^{-3}$ (solid curve) , $y=5\cdot 10^{-3}$ (dashed curve) and $y=10^{-2}$  (dotted curve). Right  panel : The $y$ dependence of  the cross section $\frac{d\sigma(e^- N \to \nu N D_s^-)}{dy\, dQ^2\, dt}$ (in pb GeV$^{-4}$) for $Q^2=0.1$ GeV$^2$, $\Delta_T = 0$  and $s=20000$ GeV$^2$.}
   \label{sigma_20000}
\end{figure}

\begin{figure}[h]
\includegraphics[width=0.5\textwidth]{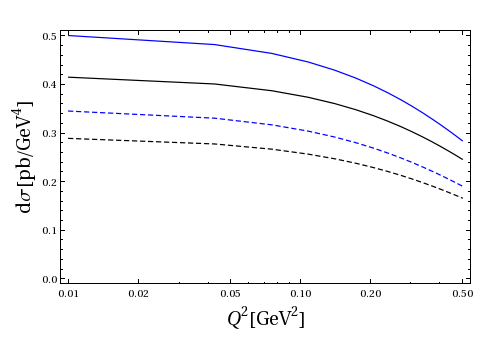}
\caption{The $Q^2$ dependence  of  the cross section $\frac{d\sigma(e^- N \to \nu N D_s^-)}{dy\, dQ^2\, dt}$ (in pb GeV$^{-4}$) for  $\Delta_T = 0$  and $s=20000$ GeV$^2$ and $y=10^{-4}$  with GK (blue lines), and simple \cite{Moutarde:2013qs}  (black lines) GPD models, and with DAs from \cite{heavyDA2} (solid lines) and \cite{Serna:2020txe}(dashed lines).}
   \label{model_dependence}
\end{figure}

Let us now present our estimates for the $D^-_s$ and $D_s^{-*}$ production cross sections. Since the gluon axial GPDs are quite smaller than the vector ones, due to the known smallness of the ratio of the relevant helicity dependent vs  spin-independent gluon PDFs $\frac{\Delta g(x)}{g(x)}$, we neglect their contributions in our following numerical analysis.

We show on Fig. \ref{sigma_821} and on Fig. \ref{sigma_20000} the $Q^2$ and $y$ dependence of the cross section for the pseudoscalar $D_s^-$ production in the low $s=820$ GeV$^2$ and high $s=20000$ GeV$^2$ energy modes of the EIC. As it may have been anticipated the $Q^2$ dependence is quite modest at small $Q^2 << M_D^2$. The $y$-dependence is quite strong resulting in the dominance of the moderate skewness region. The dependence of our results with respect to the choice of the gluon GPDs and heavy meson DAs is illustrated in  Fig. \ref{model_dependence}.

\begin{figure}[h]
\includegraphics[width=1\textwidth]{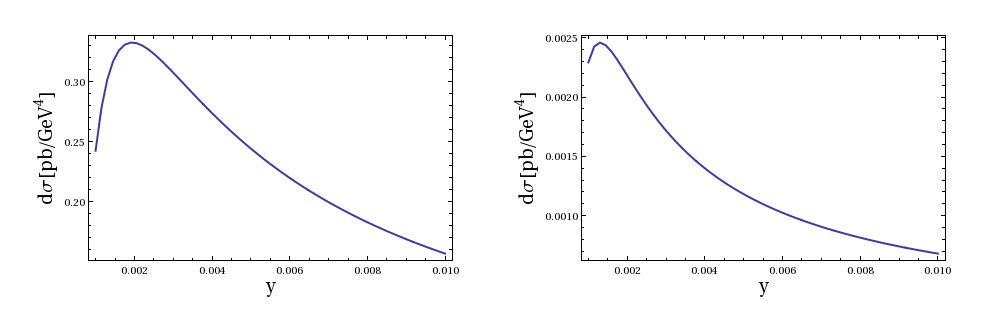}
~~~~~~~~
\caption{Left panel : The $y$ dependence  of  the longitudinal cross section $\frac{d\sigma(e^- N \to \nu N D_s^{-*})}{dy\, dQ^2\, dt}$ (in pb GeV$^{-4}$) for  $\Delta_T = 0$  and $s=20000$ GeV$^2$ and $Q^2=.1$ GeV$^2$ . Right  panel : idem for the production  cross-section of a transversely polarized $D_s^{-*}$.}
   \label{Vector}
\end{figure}

We show on Fig. \ref{Vector} both $\sigma_{00}$ and $\sigma_{--}$ for $D_s^*$ production, which could be separated by a Rosenbluth-type separation, although this separation is more difficult to perform in a charged current event than in the usual photon exchange process where the incoming energy is easier to measure in order to determine the value of $\varepsilon$. Since the $D_s^*$ vector meson is heavier than the corresponding pseudoscalar  meson, the skewness parameter $\xi$ is larger, see Eq. \ref{skewness}, and consequently the gluon GPD is  smaller and the longitudinal $D_s^*$ production cross-section is smaller than the $D_s$ one. The transverse cross-section shown on the right panel of Fig. \ref{Vector} is quite smaller than the longitudinal one.  At small $Q^2$, this can be traced back to the $Q/M_D$ additional factor present in the Dirac trace for the transversely polarized charmed meson. 

The overall conclusion is that  the cross-section is large enough for the vector charmed meson $D_s^{*-}$ to be produced through the exclusive reaction studied here, at a sizeable rate in future high luminosity  electron-ion colliders, and that it will dominantly be produced with a longitudinal polarization.  

In the case of a beam of   polarized electrons with definite helicities  only  the left-handed electrons are able to emit a $W^-$ boson, and the beam asymmetry for both the pseudoscalar and vector charmed meson production will be maximal: 
\begin{equation}
{\cal A} = \frac{d\sigma (\lambda_e = -)- d\sigma (\lambda_e = +)}{d\sigma (\lambda_e = -) + d\sigma (\lambda_e = +)}  = 1\,,
\end{equation}
which is a very clear signature of the charged current process. This may be helpful to analyze the background from neutral current (i.e. quasi real photon exchange) events with missing or misidentified  mesons in the final state.

\section{Conclusion.}
 Collinear QCD factorization has allowed us  to calculate charged current exclusive electroproduction of $D^-_s$ and $D^{*-}_s$ mesons in terms of GPDs.  Our study complements the previous calculations \cite{Siddikov:2019ahb} which were dedicated to the production of pseudoscalar and vector light mesons. The inclusive production of $D$ mesons  was also recently discussed in \cite{Siddikov:2021dfn} in the context of high multiplicity collisions at EIC.
 
 Our study also applies to the production of a $D_s^+$ or a $D^{*+}_s$ by a positron beam, with the obvious replacements of $W^-$ by $W^+$ and left-handed polarizations by right-handed ones.
 
We have demonstrated that the production cross-sections for exclusive $D_s$ charmed strange mesons, although small, are in the reach of future high luminosity electron-ion colliders making them another potential source of information for future programs aiming at the extraction of GPDs \cite{Berthou:2015oaw}. The rate for the longitudinally polarized $D_s^*$ vector meson is of the same order of magnitude as the one for the pseudoscalar $D_s$ meson. Both are in fact of the same order of magnitudes as the rates for light mesons at a $Q^2$ value of the order a few GeV$^2$ \cite{Siddikov:2019ahb}. A detailed feasibility study, taking care of the difficult reconstruction of the $D_s$ and $D_s^*$ mesons through their decay products, is needed to decide whether the reaction we study here is fully observable.


\paragraph*{Acknowledgements.}
\noindent
 We thank Benoit Blossier for useful correspondence. The work of J.W. is supported by the grant 2017/26/M/ST2/01074 of the National Science Center in Poland, whereas the work of L. S. is supported by the grant 2019/33/B/ST2/02588 of the National Science Center in Poland. This project is also co-financed by the Polish-French collaboration agreements Polonium, by the Polish National Agency for Academic Exchange and COPIN-IN2P3 and by the European Union’s Horizon 2020 research and innovation programme under grant agreement No 824093.

\end{document}